\documentclass[twocolumn,showpacs,preprintnumbers,amsmath,amssymb]{revtex4}

\usepackage{psfig}
\usepackage{graphicx}
\usepackage{dcolumn}
\usepackage{bm}

\begin{document}

\title{Non-linear exciton spin-splitting in single InAs/GaAs self-assembled quantum structures
in ultrahigh magnetic fields}

\author{A. Babinski$^{1,2}$, G. Ortner$^{3}$, S. Raymond$^{4}$, M. Potemski$^{2}$, M. Bayer$^{3}$, P. Hawrylak$^{4}$, A. Forchel$^{5}$, Z. Wasilewski$^{4}$, and S. Fafard$^{4}$}

\address{$^1$ Grenoble High Magnetic Field Laboratory, MPI/FKF and CNRS, BP166, 38042, Grenoble Cedex 9, France}
\address{$^2$ Institute of Experimental Physics, Warsaw
University, Hoza 69, 00-681 Warsaw, Poland}
\address{$^3$ Experimentelle Physik II, Universit\"at Dortmund, D-44221 Dortmund, Germany}
\address{$^4$ Institute for Microstructural Sciences, National Research Council, Ottawa, K1A 0R6, Canada}
\address{$^5$ Technische Physik, Universit\"at W\"urzburg, Am Hubland, D-97074 W\"urzburg, Germany}

\date{\today}

\begin{abstract}
We report on the magnetic field dispersion of the exciton
spin-splitting and diamagnetic shift in single InAs/GaAs quantum
dots (QDs) and dot molecules (QDMs) up to $B$ = 28 T. Only for
systems with strong geometric confinement, the dispersions can be
well described by simple field dependencies, while for dots with
weaker confinement considerable deviations are observed: most
importantly, in the high field limit the spin-splitting shows a
non-linear dependence on $B$, clearly indicating light hole
admixtures to the valence band ground state.
\end{abstract}


\maketitle

Due to its high level of component miniaturization and
integration, semiconductor nanotechnology appears to be highly
attractive for scalable quantum information processing.
\cite{Zeilinger} Semiconductor quantum dots offer charge and spin
excitations for usage as quantum bits. \cite{Loss,Imamoglu,Rossi}
Lately proposals have been made to combine the advantages that
both of them offer for these purposes: spins may provide long
coherence times, charges may offer easy coherent manipulation.
Thus electron spins could be used for information storage, but for
processing they may be swapped into charge, for example by
optically injecting electron-hole pairs through laser pulses.
\cite{CalarcoPRA03}

Proposals for quantum bit and quantum gate operation along these
lines rely heavily on well defined optical selection rules for
electron-hole excitation. Strict selection rules apply if the
valence band ground state has a pure heavy hole character. A
$\sigma^{\pm}$-polarized laser pulse then excites an electron with
momentum $S_z = \mp 1/2$ and a hole with $J_z = \pm 3/2$. This
occurs only if the dot is not yet occupied by an electron with
identical spin orientation, otherwise Pauli blocking effects would
prevent light absorption. On the basis of these selection rules
controlled injection, reliable manipulation, and accurate readout
of quantum information may be performed. However, if the hole
ground state contains light hole components, the rules become
deteriorated and the fidelity of a quantum manipulation would be
strongly reduced. Then laborious pulse shaping has to be
undertaken to reach a high enough fidelity level.
\cite{CalarcoPRA03}

Due to their high optical quality, self-assembled QDs are often
considered for the experimental realization of the proposed
schemes. The intrinsic strain induces a large splitting between
heavy and light hole, in addition to the confinement induced
splitting. Therefore for the exciton ground state 'clean'
selection rules might be expected. Experimentally, upper limits
for the light hole content can be given, for example, by making
use of the in-plane and out-of-plane polarization selectivity of
light absorption. \cite{Bastard} From such studies on flat
self-assembled In$_x$Ga$_{1-x}$As/GaAs QDs, the valence band
ground state is at least 95 \% heavy hole. \cite{CortezPRB01}
However, only a few percent light hole admixture threatens
reliable optical manipulation based on selection rules.

Here we use optical studies of single quantum structures in very
high magnetic fields up to $B$ = 28 T to obtain insight into the
angular momentum purity of the ground state exciton. Recently,
experiments on ensembles of self-assembled QDs in such high
magnetic fields have been performed. Photoluminescence spectra at
high excitation power have shown that the magnetic field
dispersion of the multiexciton emission energies can be well
described by a Fock-Darwin spectrum, renormalized by Coulomb
interactions. When two QD levels are brought into resonance by
$B$, anticrossings between the multiexciton levels occur, as the
Coulomb correlations among the confined carriers lead to mixing
and repulsion of the few particle states. In the spectra these
anticrossings were evidenced by plateaus in the field dispersion
of the inhomogeneously broadened luminescence. \cite{RaymondPRL04}

We report photoluminescence measurements of single QDs and QDMs in
magnetic fields up to 28 T applied in the Faraday-configuration.
The technique to address single quantum systems in such high
magnetic fields has only very recently been accomplished.
\cite{BabinskiPE04} For that purpose samples in which single
quantum objects were geometrically isolated (see below) have been
placed into the liquid helium insert of a cryostat at T = 4.2 K.
The cryostat is located in the bore of a Bitter-magnet with a
diameter of 50 mm. As this setup precludes optical access through
windows, a fiber system was used for the optical studies. The
laser excitation light (argon ion laser at $\lambda$ = 514.5 nm)
is led to the sample via a single-mode fiber and focussed by a
combination of two aspheric micro-lenses. The obtained spot size
is about 10 $\mu$m. The emission signal is collected by a large,
600 $\mu$m-diameter multi-mode fibre which is approached as close
as possible to the sample surface.

To address specific positions on the sample, it was mounted on a
piezo-driven stage from 'Attocube Systems'. By using such a high
stability stage that can be operated also in magnetic field,
single quantum structures can be positioned precisely below the
spatially fixed fibre optics. It allows for independent movement
in all three directions with steps down to about 10 nm at
cryogenic temperatures: first the two horizontal directions
parallel to the sample surface were adjusted, and then the
focussing of the laser spot was optimized through the
piezo-element perpendicular to the sample.  The photoluminescence
has been analyzed by a 1 m double grating monochromator and
detected by a liquid nitrogen cooled CCD-camera.

The self-assembled InAs/GaAs quantum structures studied here were
fabricated by molecular beam epitaxy. Details have been reported
elsewhere. \cite{WasilewskiJCG99} Since the focus lies on magnetic
field effects occurring on a meV energy scale, considerably
smaller than the inhomogeneous broadening, single objects had to
be isolated. This was achieved through lateral patterning of the
samples, providing mesa structures with sizes down to $\sim$ 100
nm that contained a single or a few quantum objects only. Since
these structures were arranged in a regular pattern, it was quite
easy to adjust the aforementioned setup on them. To obtain
sufficient signal, quite high optical excitation had to be used,
for which 100 mW of laser power were sent into the fibre.

\begin{figure} [h] 
  \centering
  \centerline{\psfig{figure=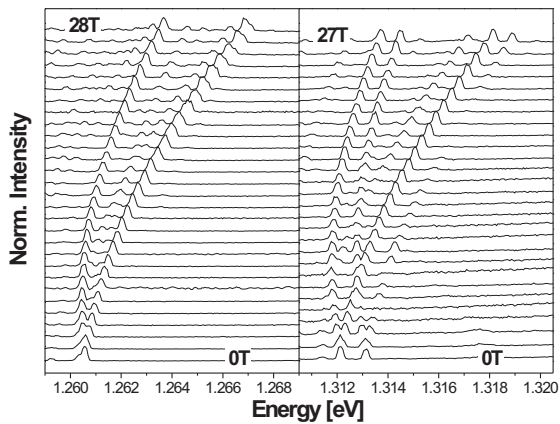,width=\columnwidth}}
  \caption{Photoluminescence spectra of single InAs/GaAs QDs
in magnetic fields up to 28 (27) T. In the left panel spectra from
a single QD emitting around 1.26 eV are shown, in the right panel
the emission occurs at about 1.31 eV.}\label{spectra}
\end{figure}

The two panels of Fig. \ref{spectra} show photoluminescence
spectra of single QDs from two different sets of structures for
varying magnetic fields. Nominally they had the same material
composition, but the ground state exciton energies are separated
by about 50 meV (also in the ensemble). In the left panel the
zero-field emission spectrum is dominated by a single line at
$\sim$ 1.26 eV, which in magnetic field splits mostly into a
doublet that can be attributed to recombination of bright excitons
with angular momentum $|M|$ = 1. \cite{BayerPRB02b} Also a few
other features with considerably weaker intensities appear in the
spectra: they can be traced to emission from other QDs (at high
$B$, for example, the features on the low energy side) or may
originate from recombination of predominantly dark excitons with
angular momentum $|M|$ = 2 that are confined in the same QD.
\cite{BayerPRB02b} Since the precise origin of these faint lines
is hard to assess we focus in the following only on the two split
features of strong intensity.

In the right panel a pair of spectral lines is observed at $B =
0$, which arise from exciton recombination at $\sim$ 1.31 eV. Two
possible origins can be foreseen for the doublet emission: either
we address two independent QDs or we observe emission from neutral
and charged excitons due to fluctuating charge occupation in a
single dot. From the present measurements, no clear decision can
be made. However, recent high field measurements show pairs of
lines with $\sim$ 1 meV separation in several mesa structures.
\cite{BabinskiPC} As it is highly unlikely to observe in a few
mesas luminescence from two QDs with virtually identical energy
separation, we exclude the first option. The attribution to
neutral and charged excitons requires identical magnetic field
dependencies, neglecting a small dot asymmetry induced exchange
splitting of less than $\sim$ 100 $\mu$eV. This is in agreement
with experiment (see below). For the dispersions of the two
features in the right panel qualitatively the same behavior is
observed as for the single line in the left panel. Again only the
doublet splittings of the two strong emission features can be
uniquely analyzed, so that we restrict to them.

\begin{figure} [h] 
  \centering
  \centerline{\psfig{figure=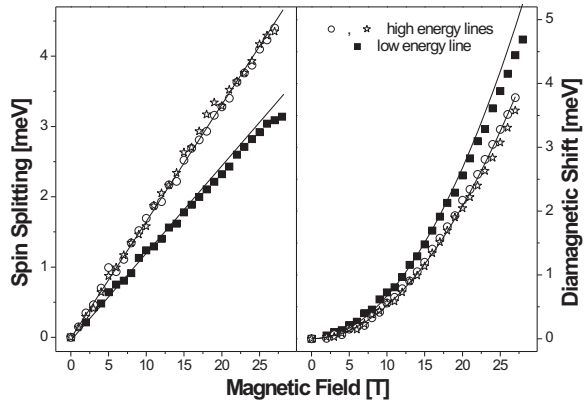,width=\columnwidth}}
  \caption{Left panel: spin-splitting of the three strongest zero field
emission lines in the spectra of Fig. \ref{spectra}. The lines are
fits to the low-field data linear in $B$. Right panel: diamagnetic
shift of the three lines. Here the solid lines are fits to the
data following $B^2$-dependencies. Data of the low (the solid
symbols) and high energy (the open symbols) emitting dots are
shown.}\label{qdspindia}
\end{figure}

From the data in Fig. \ref{spectra} we have extracted the
spin-splitting of the emission lines as well as the diamagnetic
shift of the split line centers. Fig. \ref{qdspindia} shows the
magnetic field dependencies of these quantities. Based on the
results of previous studies up to about 10 T \cite{diamagspin},
two features are expected for these dispersions in the low field
limit: (a) The diamagnetic shift should depend quadratically on
magnetic field, $\Delta_{diamag} = \frac{e^2}{8} \left(
\frac{\langle x_e^2 + y_e^2 \rangle}{m_e} + \frac{\langle x_h^2 +
y_h^2 \rangle}{m_h} \right) B^2$, as we address quantum structures
in the strong confinement regime, in which the quantization energy
is considerably larger than the electron-hole interaction energy.
Here we assume that the magnetic field points along the
$z$-direction. (b) the exciton spin-splitting should vary linearly
with magnetic field $\Delta_{spin} = \left( g_e + g_h \right)
\mu_B B$, where the $g_i$ are the g-factors of electron and hole.

These expectations are fulfilled in both cases in the range of low
$B$-fields. For the dots emitting at higher energies the linear
and quadratic field dispersions for $\Delta_{spin}$ and
$\Delta_{diamag}$, respectively, even extend towards the highest
magnetic fields, as corresponding fits to the data show (the solid
lines). In contrast, for the dot emitting at lower energies
deviations are observed: The diamagnetic shift can no longer be
described by a pure $B^2$-form, but a satisfactory fit can be
obtained only by inclusion of a contribution that goes linearly
with magnetic field. Also for the spin-splitting deviations from
the linear field dependence are observed above 20 T.

Let us comment first on the diamagnetic shift: the expression
given above can be used to estimate the extension of the electron
and hole wave functions in the dot plane. As the two dot types
have nominally the same material compositions, we assume the same
carrier masses for them. The difference in their emission energies
has to be attributed then to different dot sizes with the size of
the high energy emitting dots being smaller, leading to larger
quantization energies. This is supported by the diamagnetic shift
data, which is smaller for this dot than for the low energy
emitting QD. \cite{biexciton}

However, the expression for $\Delta_{diamag}$ has been derived by
assuming that the magnetic field induced confinement is clearly
weaker than the geometric one, so that the $B$-field effects can
be treated by perturbation theory. \cite{diamagth} The confinement
strengths are characterized by two energy scales: the energy
splitting between the confined QD shells is a measure for the
geometric confinement, the magnetic confinement is given by the
cyclotron energy. For the dots under study the splitting between
p- and s-shell emission is about 50 meV for the low energy QD,
while it is $\sim$ 70 meV for the high energy QD, in accord with
their different dot sizes. These values have been taken from high
excitation state-filling spectroscopy. They are largely determined
by the splitting of the confined electron levels due to its small
mass. These energies have to be compared to the cyclotron energy
$\hbar \omega_c = \hbar \left( e B/ m \right)$. Assuming an
electron mass of 0.04 in the intermixed InGaAs dot material, the
corresponding cyclotron energy will exceed 50 meV above 20 T. This
comparison demonstrates that perturbation theory can no longer be
used in this field range, as the magnetic confinement even becomes
dominant there for the low energy QD. This will naturally lead to
linear field contributions in the energy dispersion, because the
level structure approaches Landau-level like behavior. In
contrast, for the high energy dots a pure quadratic field
dependence still provides a reasonable description of the data.

Now let us turn to the exciton spin-splitting as function of
magnetic field. Non-linearities of its $B$-dependence are well
known for systems of higher dimensionality such as quantum wells.
\cite{2D} A contribution arising from possible field-dependence of
the electronic g-factor is likely to be neglected, and the
non-linearities mostly originate from heavy hole-light hole mixing
which varies with magnetic field. \cite{KotlyarPRB01} Even though
a quantitative understanding can be obtained only on base of
detailed band structure calculations \cite{bandstructureth} (which
are beyond the scope of this paper), the spin-splitting thereby
permits to take insight into valence band mixing. For the dot with
strong confinement the $B$-linear splitting over the whole field
range indicates that the hole s-shell states have indeed pure
heavy-hole character within the experimental accuracy. For the
less strongly confined dot, on the other hand, the non-linear
dependence at high fields indicates light hole admixtures which
aggravates their use in quantum information schemes.

\begin{figure} [h] 
  \centering
  \centerline{\psfig{figure=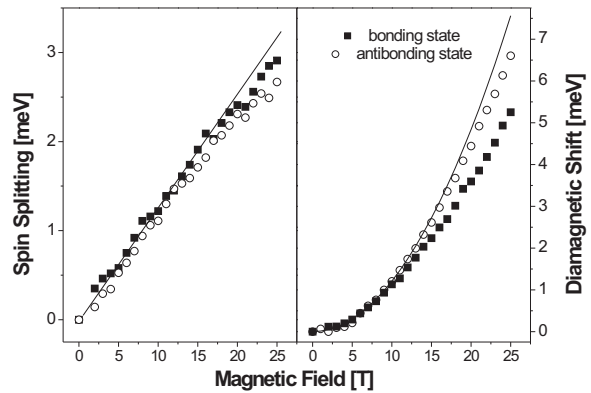,width=\columnwidth}}
  \caption{Left panel: Spin splitting of the 'bonding' (the full
  symbols) and the 'antibonding' (the open symbols) s-shell exciton
  states in an InAs/GaAs QDM of high symmetry. The barrier width
  was 7 nm. The solid line is a fit to the low field data using a $B$-linear form.
  The right panel gives the corresponding diamagnetic shifts.
  The solid line is a fit proportional to $B^2$ to the low field data for the antibonding state.
\cite{OrtnerPRB05}}\label{qdmspindia}
\end{figure}

To obtain further insight, we have also studied quantum dot
molecules (QDMs) in such high magnetic fields. The dots forming
the molecules are separated from each other by a few nm wide
barrier and are nominally identical to the low energy emitting QDs
discussed above. The center of emission energies from the
'bonding' and the 'antibonding' exciton states is as well located
in this energy range. \cite{OrtnerPRB05} Here we focus on
structures with a simple fine structure splitting: the emission
spectrum exhibits at zero field a single emission line both for
the bonding and antibonding states, which show a doublet
splitting, that can easily be traced in magnetic field.

The solid symbols in the right panel of Fig. \ref{qdmspindia} show
the diamagnetic shift of the emission from the 'bonding' exciton
state versus $B$ for a molecule with a 7 nm barrier. It is larger
than observed for the QD sample and exceeds 5 meV at 28 T. Also
here the field dispersion cannot be well described by a simple
$B^2$ dependence, but linear terms have to be included for the
same reasons as in the dot case, since also here the splitting
between the p- and the s-shell features is about 50 meV. The same
holds for the shift of the 'antibonding' state (the open symbols),
which is considerably larger than that of the 'bonding' state. At
28 T it reaches almost 7 meV.

Let us turn to the field dispersion of the spin-splitting of the
two features, shown in the left panel of Fig. 3. Again solid and
open symbols give the data for 'bonding' and 'antibonding' states,
respectively. For both of them, pronounced deviations from a
$B$-linear dispersion are observed, which are more prominent than
for the QDs, as evidenced also by their emergence at smaller
fields.

Electronically coupled QDs represent an architecture that could be
used for coupling and entangling the states of two quantum bits,
each confined in one of the dot structures. Control of the
entanglement could be provided by application of an electric field
along the molecule axis or by a barrier of tunable height between
the dots. For optically controlled schemes the influences arising
from valence band mixing are therefore equally important as for
single QDs. From the data we conclude that the QDM structures
available for the present high field studies do not seem to be
very well suited for such quantum information purposes as the band
mixing induced non-linearity of the spin-splitting dependence on
$B$ is enhanced by the molecule formation.

In summary, we have shown that the valence band ground state in
quantum structures with comparatively weak confinement shows light
hole admixtures in magnetic field, lifting the strict selection
rules for the exciton states in which they are involved. However,
by a proper tailoring of the quantum confinement valence band
mixing can be reduced to an extent, that it appears to be no
longer relevant for coherent manipulation of pure angular momentum
states. On the other hand, increase of the confinement might lead
to other problems such as significant exciton dephasing through
acoustic phonon related transitions, leading to a broad background
in the dot spectrum. \cite{BorriPRL01}

Support by the Deutsche Forschungsgemeinschaft (research program
'Quantum Optics in Semiconductor Nanostructures' and GK 726), by
the BMBF 'nanoquit' research program and by the NRC-Helmholtz
Joint Research program is gratefully acknowledged. G.O. also
thanks the EC for support (grant SSATA EC Program No.
RITA-CT-2003-505474).


\begin{references}
\bibitem{biexciton}
The different dot sizes are also confirmed by data on the
biexciton binding energy. For the low energy QD it is $\sim$ 3
meV, while for the high energy QD it is 4.2 meV, as expected from
the increased quantum confinement.
\end{references}
\end{document}